\documentclass[a4paper,onecolumn,11pt,accepted=2020-07-28]{quantumarticle}
\pdfoutput=1
\usepackage[utf8]{inputenc}
\usepackage[english]{babel}
\usepackage[T1]{fontenc}
\usepackage{amsmath,amssymb}
\usepackage{soul}
\usepackage{hyperref}
\usepackage[numbers,compress]{natbib}
\usepackage{tikz}
\usepackage{lipsum}

\newcommand{\Tr}{\mathop{\mathrm{Tr}} \nolimits}

\newcommand{\op}[1]{\hat{#1}}
\newcommand*\pFqskip{8mu}
\catcode`,\active
\newcommand*\pFq{\begingroup
	\catcode`\,\active
	\def ,{\mskip\pFqskip\relax}%
	\dopFq
}
\catcode`\,12
\def\dopFq#1#2#3#4#5{%
	{}_{#1}F_{#2}\biggl(\begin{array}{@{ }c@{ }}#3\\#4\end{array};#5\biggr)%
	\endgroup
      }
      
\begin{document}

\title{Wigner function for SU(1,1)} 

\author{U.~Seyfarth}
\affiliation{Max-Planck-Institut f\"ur die Physik des Lichts,
  Staudtstra{\ss}e 2, 91058~Erlangen, Germany}
\orcid{0000-0001-8657-4272}

\author{A.~B.~Klimov}
\affiliation{Departamento de F\'{\i}sica, 
Universidad de Guadalajara, 44420~Guadalajara, 
Jalisco, Mexico}
\orcid{0000-0001-8493-721X}

\author{H. de Guise}
\affiliation{Department of Physics, Lakehead University,
  Thunder Bay, Ontario P7B 5E1, Canada}
\orcid{0000-0002-1904-4287Mi}

\author{G.~Leuchs}
\affiliation{Max-Planck-Institut f\"ur die Physik des Lichts,
  Staudtstra{\ss}e 2, 91058~Erlangen, Germany}
  \affiliation{Institute for Applied Physics,Russian Academy of Sciences, 630950 Nizhny Novgorod, Russia}
  \orcid{0000-0003-1967-2766}

\author{L.~L.~S\'{a}nchez-Soto}
\affiliation{Max-Planck-Institut f\"ur die Physik des Lichts,
  Staudtstra{\ss}e 2, 91058~Erlangen, Germany}
\affiliation{Departamento de \'Optica,
Facultad de F\'{\i}sica, Universidad Complutense,
28040~Madrid, Spain}
\orcid{0000-0002-7441-8632}

\begin{abstract}
  In spite of their potential usefulness, Wigner functions for systems
  with SU(1,1) {symmetry} have not been explored thus far.  We
  address this problem from a physically-motivated perspective, with
  an eye towards applications in modern metrology. Starting from two
  independent modes, and after getting rid of the irrelevant degrees
  of freedom, we derive in a consistent way a Wigner distribution for
  SU(1,1).  This distribution appears as the expectation value of the
  displaced parity operator, which suggests a direct way to
  experimentally sample it. We show how this formalism works in some
  relevant examples.

  \medskip

 \noindent \textbf{Dedication}: While this manuscript was under review, we learnt with great sadness of the untimely passing of our colleague and friend Jonathan Dowling.  Through his outstanding scientific work, his kind attitude, and his inimitable humor, he leaves behind a rich legacy for all of us. Our work on SU(1,1) came as a result of long conversations during his frequent visits to Erlangen. We dedicate this paper to his memory.
\end{abstract}

\maketitle

\section{Introduction}

Phase-space methods represent a self-standing alternative to the conventional Hilbert-space formalism of quantum theory.  In this approach, observables are $c$-number functions instead of operators, with the same interpretation as their classical counterparts, although composed in novel algebraic ways. Quantum mechanics thus appears as a statistical theory on phase space, which {can make} the corresponding classical limit emerge in a natural and intuitive manner.

The realm of the method was established in the groundbreaking {work} of {Weyl~\cite{Weyl:1927aa} and} Wigner~\cite{Wigner:1932uq}. Later, Groenewold~\cite{Groenewold:1946aa} and Moyal~\cite{Moyal:1949fk} established a solid {foundation} that has developed over time into a complete discipline {useful in many} diverse fields~\cite{Schroek:1996fv,Schleich:2001hc,QMPS:2005mi,Weinbub:2018aa}.

The main ingredient of this approach is a \emph{bona fide} mapping that relates operators to functions defined on a smooth manifold, endowed with a very precise mathematical structure~\cite{Kirillov:2004aa}.  However, this mapping is not unique: actually, a whole family of functions can be consistently assigned to each operator. In particular, quasiprobability distributions are the functions connected with the density operator~\cite{Tatarskii:1983uq,Balazs:1984cr,Hillery:1984oq,Lee:1995tg}. For {continuous variables, such as Cartesian position and momentum, the quintessential example that fuelled the interest for this field,} the most common choices are the $P$ (Glauber-Sudarshan)~\cite{Glauber:1963aa,Sudarshan:1963aa}, $W$~(Wigner)~\cite{Wigner:1932uq}, and  $Q$~(Husimi)~\cite{Husimi:1940aa} functions, respectively.

This formalism has been applied to other different {dynamical
groups}~\footnote{We adhere to the usual convention that a Lie group $G$ (with Lie algebra $\mathfrak{g}$) is a dynamical group if the Hamiltonian of the system under consideration can be expressed in terms of the generators of $G$ (that is, the element of the Lie algebra $\mathfrak{g}$)~\cite{Zhang:1990aa}.}. Probably, the most widespread {example} beyond the harmonic oscillator is that of SU(2), with the Bloch sphere as associated phase space~\cite{Stratonovich:1956aa,Berezin:1975aa,Varilly:1989bh}; this {case} is of paramount importance in dealing with spinlike systems~\cite{Agarwal:1981zr, Dowling:1994sw,Nieto:1998cr,Heiss:2000aa,Chumakov:2000aa,Klimov:2017aa}. Suitable results have also been found for the Euclidean group E(2), this time with the cylinder as phase space~\cite{Mukunda:1979uq,Plebanski:2000fk,Rigas:2011by,Kastrup:2016aa}; this is of primary importance in treating the orbital angular momentum of twisted photons~\cite{Molina:2007kn,Franke-Arnold:2008sw}.  {Additional applications to more general dynamical groups have also appeared in the literature~\cite{Brif:1998qf,Mukunda:2005aa,Klimov:2010aa,Tilma:2016aa}. Moreover,} the basic notions have been successfully extended to discrete qudits, where {the} phase space is a finite grid~\cite{Wootters:1987kl,Galetti:1988ff,Galetti:1992pi,Gibbons:2004ye,Vourdas:2007tw,Bjork:2008ab,Klimov:2009aa}.

Surprisingly, the phase-space description of systems having SU(1,1) symmetry has received comparatively little attention~\cite{Orowski:1990aa,Alonso:2002aa}, in part because the representation theory of this group is not as familiar as SU(2) or even E(2).  However, SU(1,1) plays a major role in connection with what can be called two-photon effects~\cite{Wodkiewicz:1985aa,Gerry:1985aa,Gerry:1991aa,Gerry:1995kq}. The topic is experiencing a revival in popularity due to the recent realization of a nonlinear SU(1,1) interferometer~\cite{Jing:2011aa, Hudelist:2014aa}.  According to the {pioneering proposal of} Yurke \textit{et al.}~\cite{Yurke:1986yg}, this device would allow one to improve the phase measurement sensitivity in a stunning manner~\cite{Chekhova:2016aa}.

In spite of the importance of these systems, the mathematical complexity of the group SU(1,1)~\cite{Novaes:2004aa} has prevented a proper phase-space description. In this paper we approach this question {resorting to a} physics-based approach. For SU(2), one can model the description in terms of a superposition of two harmonic modes. In technical terms, this corresponds to the Jordan-Schwinger bosonic realization of the algebra $\mathfrak{su}(2)$~\cite{Chaturvedi:2006kl}. Here, we propose a similar way to deal with $\mathfrak{su}(1,1)$: starting with two orthonormal modes, and using the standard tools for continuous variables, we eliminate the spurious degrees of freedom and we get a description {on the upper sheet of a two-sheeted hyperboloid}, which is the natural arena to represent the physics associated to these systems.

Our final upshot is that the Wigner function can be expressed as the average value of the displaced parity operator. This is reassuring, for it is also the case for continuous variables~\cite{Royer:1977aa}. Moreover, as this property has been employed for the direct sampling of the Wigner function for a quantum field~\cite{Banaszek:1999aa,Bertet:2002aa,Harder:2016aa}, our result opens the way for the experimental determination of the Wigner function for SU(1,1).    
  
\section{Phase-space representation of a single mode}

To keep the discussion as self-contained as possible, we first briefly summarize the essential ingredients of phase-space functions for a harmonic oscillator that we shall need for our purposes. 

We consider the standard oscillator described by annihilation and creation operators $\op{a}$ and $\op{a}^{\dagger}$, which obey the bosonic  commutation relation 
\begin{equation}
  \label{eq:ccr}
  [\op{a}, \op{a}^{\dagger} ] = \op{\openone} \,  .
\end{equation}
They are the generators of the Heisenberg-Weyl algebra, which has become the hallmark of noncommutativity in quantum theory~\cite{Binz:2008oq}. The classical phase space is here the complex plane $\mathbb{C}$.

These complex amplitudes are expressed in terms of the quadrature operators $\op{x}$ and $\op{p}$ as 
\begin{equation}
  \label{eq:aadag}
  \op{a} = \frac{1}{\sqrt{2}} (\op{x} + i \op{p} )  \, ,
  \qquad \qquad
  \op{a}^{\dagger} = \frac{1}{\sqrt{2}} (\op{x} - i \op{p} ) \, ,
\end{equation}
and the commutation relation (\ref{eq:ccr}) reduces then to the canonical form $ [\op{x}, \op{p} ] = i \, \op{\openone}$.

A central role in what follows will be played by the unitary operator
\begin{equation}
  \op{D} (\alpha) =
  \exp( \alpha \op{a}^{\dagger} - \alpha^{\ast} \op{a} ) \, ,
  \qquad 
  {\alpha \in \mathbb{C} \, ,}
\end{equation}
which is called the displacement operator for it displaces a state localized in phase space at $\alpha_{0}$ to the point $\alpha_{0}+\alpha$.  The Fourier transform of the displacement is the Cahill-Glauber kernel~\cite{Cahill:1969aa}
\begin{equation}
  \label{wCG}
  \op{w} (\alpha) = \frac{1}{\pi^2}
  \int_{\mathbb{C}} 
  \,\exp(\alpha \beta^\ast - \alpha^\ast \beta)
  \, \op{D} (\beta) \, d \beta \, ,
\end{equation}
which is an instance of a Wigner-Weyl quantizer~\cite{Case:2008aa}.

The operators $\op{w} (\alpha)$ constitute a complete trace-orthonormal set that transforms properly under displacements; {that is}
\begin{equation}
  \label{eq:HWKernelDisp}
  \op{w} (\alpha) = 
  \op{D} (\alpha) \, \op{w} (0) \, \op{D}^\dagger (\alpha ) 
  = 2 \;  \op{D} (\alpha)  \,  (-1)^{\op{a}^{\dagger} \op{a}} \, 
  \op{D}^\dagger (\alpha) \, ,  
\end{equation}
where $\op{w}(0)=\int_{\mathbb{C}} \op{D}( \beta ) \, d\beta  = 2  \op{P}$,
and 
 \begin{equation}
  \label{eq:2}
  \op{P} = (-1 )^{\op{a}^{\dagger} \op{a}}  \, .
\end{equation}
{In this way, $\op{w} (\alpha)$ appear as the displaced parity operator~\cite{Royer:1977aa}.}

If $\op{A}$ is an arbitrary (trace-class) operator acting on the Hilbert space of the system, the Wigner-Weyl quantizer allows one to associate to $\op{A}$ a function $W_{\op{A}}( \alpha)$ representing the action of the corresponding dynamical variable in phase space:
\begin{equation}
  \label{eq:Asy}
  W_{\op{A}} (\alpha )  =   \Tr [ \op{A} \,\op{w} (\alpha ) ] \, . 
\end{equation}
The function $W_{\op{A}}(\alpha)$ is the symbol of the operator $\op{A}$. Such a map is one-to-one, so we can invert it to get the operator from its symbol through
\begin{equation}
  \label{eq:WigWeyl}
  \op{A} = \frac{1}{(2\pi)^{2}} \int_{\mathbb{C}} 
  \op{w} (\alpha ) \, W_{\op{A}}(\alpha ) \, \, d\alpha \, .
\end{equation}

We focus on what follows on the Wigner function, although {the discussion}  can be immediately extended to any other quasiprobability. Actually, the Wigner function  is nothing but the symbol of the density matrix $\op{\varrho}$.  Consequently,  
\begin{eqnarray}
  \label{eq:Wigcan}
  &  W_{\op{\varrho}}(\alpha )  =   \Tr [ \op{\varrho} \,\op{w} (\alpha) ]
  \, , &  \nonumber \\ 
  & &  \label{eq:HWWignerDef} \\
  & \op{\varrho}   =   \displaystyle 
  \frac{1}{(2\pi)^{2}} \int_{\mathbb{C}} \op{w}(\alpha)
  W_{\op{\varrho}} (\alpha ) \, d\alpha \, . &
  \nonumber
\end{eqnarray}

The $W_{\op{\varrho}}(\alpha )$ defined in (\ref{eq:HWWignerDef}) fulfills the basic properties required for any good probabilistic description~\cite{Brif:1998qf}.  First, due to the Hermiticity of $\op{w}( \alpha) $, it is real for Hermitian operators. Second, the probability distributions for the canonical variables can be obtained as the corresponding marginals.  Third, $W_{\op{\varrho}} (\alpha )$ is translationally covariant, which means that for the displaced state $\op{\varrho}^\prime = \op{D}(\alpha^{\prime}) \, \op{\varrho} \, \op{D}^\dagger (\alpha^{\prime})$, one has 
\begin{equation}
  \label{eq:HWProps3}
  W_{\op{\varrho}^\prime} (\alpha) = W_{\op{\varrho}} (\alpha
  -\alpha^{\prime} ) \, ,
\end{equation}
so that {the Wigner function follows} displacements rigidly, without changing its form, reflecting the fact that physics should not depend on {any choice} of the origin. 

Finally, the overlap of two density operators is proportional to the integral of the associated Wigner functions
\begin{equation}
  \label{eq:HWProps4}
  \Tr ( \op{\varrho} \,\op{\varrho}^{\prime} ) \propto
  \int_{\mathbb{C}} W_{\op{\varrho}} (\alpha )
  W_{\op{\varrho}^{\prime}} (\alpha )  \, d\alpha  \, ,
\end{equation}
{provided the integral converges.} This property (known as traciality) offers practical advantages, {because it allows} one to predict the statistics of any outcome, once the Wigner function of the measured state is known.

To conclude, we mention that the displacements {also constitute} a basic ingredient in the concept of coherent states. If we choose a fixed normalized reference state $ | \Psi_{0}\rangle $, we have~\cite{Perelomov:1986kl}
\begin{equation}
  | \alpha \rangle = \op{D} (\alpha) \, | \Psi_{0} \rangle \, ,  
  \label{eq:defCS}
\end{equation}
so they are parametrized by phase-space points. These states have a number of remarkable properties inherited from those of $\op{D} (\alpha)$.  The standard choice for the fiducial vector $| \Psi_{0} \rangle$ is the vacuum $| 0 \rangle $ {(or, more generally, a highest or lowest weight state).}

\section{Phase-space representation of two modes}

Next, we consider the superposition of two modes in two orthogonal directions, say $x$ and $y$, with momenta $p_{x}$ and $p_{y}$, respectively. Since they are kinematically independent, the complex amplitudes of these modes (denoted by $\op{a}$ and $\op{b}$) commute  ($[\op{a}, \op{b} ] = 0$) and the total Wigner-Weyl quantizer can be expressed as the product of the corresponding ones for each mode:
\begin{equation}
  \op{w} (\alpha, \beta ) =  \op{w} (\alpha) \,
  \op{w} ( \beta ) \, .
  \end{equation}
{With the form given in Eq.~(\ref{eq:HWKernelDisp})} and disentangling the exponentials, we get
\begin{equation}
  \op{w} (\alpha, \beta )  =   4
  \exp [ - 2 ( |\alpha |^{2} - |\beta |^{2}) ]
 ( - 1)^{\op{a}^{\dagger} \op{a} + \op{b}^{\dagger} \op{b}} 
  \exp [ - 2 ( \alpha \op{a}^{\dagger}  -  \beta^{\ast} \op{b} ) ] \,
 \exp [  2 ( \alpha^{\ast} \op{a}  - \beta \op{b}^{\dagger} ) ] \, .        
\end{equation}
As we can see, this kernel depends on the four real variables $\alpha = (x ,p_{x})$ and $\beta = (y ,p_{y})$. As a consequence, the resulting Wigner function $W (\alpha , \beta )$ contains all the {information on the two modes}, but it is hard to grasp any physical flavor from it: in particular, it cannot be plotted, which is always a big advantage in depicting complex phenomena. To avoid this drawback we use the
parametrization 
\begin{equation}
  \label{eq:Hopf}
  \alpha = r e^{i (\chi +  \varphi)/2} \cosh (\tau/2) \, ,
  \qquad
  \beta = r e^{i ( \chi -  \varphi)/2}  \sinh (\tau/2) \, ,
\end{equation}
where the radial variable $r^{2} = |\alpha|^{2} - |\beta|^{2}$ represents the
difference in intensities between the two modes. {We can safely take $|\alpha | > |\beta|$, for the opposite case can be obtained by just a relabelling of modes, with no physical consequences. The} parameters $\chi$ and $\tau$ can be interpreted as azimuthal and ``polar'' angles on a two-sheeted hyperboloid $\mathbb{H}_{2}$~\cite{Hasebe:2019aa}. A similar parametrization as in (\ref{eq:Hopf}), wherein the hyperbolic functions are replaced with trigonometric ones, maps two complex modes into the Bloch sphere $\mathbb{S}_{2}$. This is an instance of a Hopf fibration~\cite{Mosseri:2001aa}. Therefore, the hyperboloid $\mathbb{H}_{2}$ can be properly called the Bloch hyperboloid and the map \eqref{eq:Hopf} is  a  noncompact Hopf fibration~\cite{Hasebe:2010aa}. 

 After some lengthy algebra, the kernel can be recast in the form
 \begin{equation}
  \op{w} (r, \chi, \tau, \varphi  )  = 
 4\exp ( -2r^{2} )
   \left( -1\right) ^{\op{N}} 
 \op{S} ( \zeta ) \;
 \exp  ( -2re^{i\chi }e^{i\varphi}\op{a}^{\dagger} )
 \exp  ( 2re^{-i\chi }e^{-i\varphi }\op{a} ) \,
 \op{S}^{\dagger} (\zeta) \, ,
   \end{equation}
where $\op{N} = \op{a}^{\dagger}\op{a} + \op{b}^{\dagger} \op{b}$ is the total number and we have introduced the two-mode squeeze operator 
\begin{equation}
\label{eq:squeezeop}
  \op{S} ( \zeta) =
  \exp( \zeta \op{K}_{+} - \zeta^{\ast}  \op{K}_{-} ) \, ,
\end{equation}
{with $\zeta = \tfrac{1}{2} \tau e^{ i \chi}$. This operator is defined in terms of} the two-mode realization of the $\mathfrak{su}(1,1)$ algebra 
\begin{equation}
  \op{K}_{+}= \op{a}^\dagger \op{b}^\dagger \,  ,
  \qquad
  \op{K}_{-}= \op{a}\op{b} \, ,
  \qquad
  \op{K}_{0}=\tfrac{1}{2}(\op{a}^\dagger \op{a} +
  \op{b}^\dagger \op{b} + \openone )\, ,
\end{equation}
with commutation relations
\begin{equation}
  [ \op{K}_{0}, \op{K}_{\pm} ] = \pm \op{K}_{\pm}\, ,
  \qquad
  [\op{K}_{-}, \op{K}_{+}] = 2 \op{K}_{0} \, .
  \end{equation}
Using the {Baker-Campbell-Hausdorff} formula, one can check that 
\begin{eqnarray}
  \op{S} (\zeta ) \, \op{a} \, \op{S}^{\dagger}(\zeta ) = 
  \op{a} \cosh |\zeta| -
  \op{b}^{\dagger} \, e^{i \arg  \zeta}\sinh |\zeta | \, ,
\end{eqnarray}

{Notice that the transformation $\op{S} (\zeta)$ depends only on the sum of the phases $e^{i \chi}$. This makes the phase $\varphi$ irrelevant and, consequenly, we proceed to integrate over $\varphi$ to get} 
\begin{eqnarray}
\op{w} (r, \zeta ) & = & \frac{1}{2\pi}\int_{-\pi}^{\pi}d\varphi \
\op{w} (r, \chi, \tau, \varphi) \nonumber  \\
& = & 4 \exp ( -2r^{2} ) (-1)^{\op{N}} \op{S}(\zeta )\
      \sum_{k=0}^{\infty} \frac{(2r)^{2k}(-1)^{k}}{k!^{2}}
      \op{a}^{\dagger k}\op{a}^{k}\ \op{S}^{\dagger}(\zeta ) \,.  \
\end{eqnarray}
Finally, we integrate over $r$, {which is tantamount to averaging over intensity information:}
\begin{equation}
  \op{w} (\zeta )   =  2 \int_{0}^{\infty} \ dr \,r \, 
  \op{w} (r,\zeta)  =   2 ( -1 )^{\op{N}} 
  \op{S}(\zeta ) \,\sum_{k=0}^{\infty} \frac{(-2)^{k}}{k!}
  \op{a}^{\dagger k}\op{a}^{k} \; \op{S}^{\dagger}(\zeta ) \,.
\end{equation}
If we {realize that} 
\begin{equation}
  \sum_{k=0}^{\infty}\frac{z^{k}}{k!}\op{a}^{\dagger k}\op{a}^{k} =
  (z+1)^{ \op{a}^{\dagger}\op{a}}\,,
\end{equation}
we arrive at our central result
\begin{equation}
  \boxed{\op{w} (\zeta ) = 2 \op{S} (\zeta )  \,  ( -1)^{\op{K}_{0}}  \,
  \op{S}^{\dagger}(\zeta )\,}
\end{equation}
Since $(-1)^{\op{K}_{0}}$ is the SU(1,1) parity and $\op{S} (\zeta)$ is a displacement operator, this shows that the Wigner function for SU(1,1) can be understood much in the same way as for the harmonic oscillator: just a displaced parity. 

{The optical parity operator have been considered as a candidate for approaching the highest level of sensitivity in the detection of small phase shifts via optical interferometry~\cite{Gerry:2010aa}. This idea is adaption of a proposal by Bollinger \emph{et al.}~\cite{Bollinger:1996aa} in the context of spectroscopy for a collection of maximally entangled two-level trapped ions, in which parity is determined via counting the number of ions of the sample that populate the excited state. This detection has been} recently proposed as a scheme to beat the Heisenberg limit in SU(1,1) interferometry~\cite{Anisimov:2010aa,Plick:2010aa,Gerry:2011aa,Li:2016aa}. Bear in mind though that the SU(1,1) parity is not, in general, the parity of the photon numbers~\cite{Hach:2018aa}.

The operator $\op{S}(\zeta)$ displaces by a complex number $\zeta \in \mathbb{C}$. Moreover, there is a one-to-one correspondence between $\zeta \in \mathbb{C}$    and the upper sheet of the hyperboloid, {usually denoted by $\mathbb{H}_{2}$: it is} established via stereographic projection from the south pole, so that
\begin{equation}
  \xi = \tanh (\tau/2) e^{ i \chi} \Leftrightarrow
  \mathbf{n} = (\cosh \tau, \sinh \tau \cos \chi, \sinh \tau \sin
  \chi)\, ,
  \end{equation}
where $\mathbf{n}$ is a unit vector on $\mathbb{H}_{2}$, with the  metric $\mathbf{n}^{2}= n_{0}^{2} - n_{1}^{2} - n_{2}^{2}$. {Note that $\zeta$ are points in $\mathbb{C}$, whereas $\xi$ are points in $\mathbb{H}_{2}$, but both are equivalent.}  This construction provides a complex structure on the upper sheet of  the hyperboloid $\mathbb{H}_{2}$, which can be treated as a noncompact complex manifold.
  
\section{Explicit form of the Wigner function for SU(1,1)}

To gain further insights into this formalism, we will {obtain} the
structure of the Wigner function for SU(1,1) {in more details.}  

Before going ahead we recall that the irreducible representations (irreps) of SU(1,1) are labeled by the eigenvalues of the Casimir operator
\begin{equation}
  \op{K}^{2} = \op{K}_{0}^{2} - \op{K}_{1}^{2} -
  \op{K}_{2}^{2}  = k (k-1) \openone \, ,
\end{equation}
where $\op{K}_{\pm} = \pm i ( \op{K}_{1} \pm i \op{K}_{2} )$. The  irrep $k$ is carried by a Hilbert space spanned by the common eigenstates of $\op{K}^{2}$and $\op{K}_{0}$: $\{ |k, \mu \rangle\,  :  \mu = k, k+1, \ldots \}$. All unitary irreps are infinite dimensional. There are several different series of irreps for SU(1,1) fixed by the domains of the eigenvalues $k$~\cite{Bargmann:1947fk}.  For representations in the positive discrete series, where $2k=1,2,3,\ldots$ and including the two limit of discrete series with $k=1/4$ and $3/4$, the action of the generators $\{ \op{K}_{0} , \op{K}_{\pm}\}$ therein is
\begin{eqnarray}
  \op{K}_{0}  |k, \mu \rangle &= & \mu |k, \mu \rangle \nonumber \\
\\
  \op{K}_{\pm}  |k, \mu \rangle & = & \sqrt{(\mu \pm k) (\mu \mp k \pm 1)}\,
  |k, \mu \pm 1 \rangle \, . \nonumber 
\end{eqnarray}
This carrier space is denoted by~$\mathcal{D}_{k}^{+}$.

If the number of excitations in modes $a$ and $b$ are $n_{a}$ and
$n_{b}$, respectively, then $k$ and $\mu$ satisfy 
\begin{equation}
  k= \tfrac{1}{2} (|n_{a} - n_{b}| + 1) \, ,
  \qquad
  \mu = \tfrac{1}{2} (n_{a} + n_{b} +1) \, .
\end{equation}
{As discussed before, $n_{b} > n_{a}$  can be obtained from $n_{a} > n_{b}$ by just a relabelling of modes, with no physical consequences. Therefore , we consider $\pm (n_{a} - n_{b})$ to be equivalent irreps.}  The total Hilbert space of the two oscillators decomposes then  as
\begin{equation}
  \label{eq:split}
\mathcal{H}_{a} \otimes \mathcal{H}_{b} =
\mathcal{D}_{\frac{1}{2}}^{+}
\oplus
\mathcal{D}_{1}^{+}
\oplus
\mathcal{D}_{\frac{3}{2}}^{+} \oplus \cdots .
\end{equation}
This decomposition allows us to expand any (pure) state in an
SU(1,1)-invariant way; viz,
\begin{equation}
  |\Psi \rangle = \sum_{k} \sum_{\mu} \Psi_{k\mu} |k, \mu \rangle \, ,
\end{equation}
where $\Psi_{k\mu} = \langle k, \mu | \Psi \rangle$. The Wigner function reads
\begin{equation}
  \label{eq:finalds}
  W(\zeta)  =  \langle \Psi | \op{w} (\zeta ) | \Psi \rangle
   =   \sum_{k} \sum_{\mu, \mu^{\prime}} \Psi_{k \mu}^{\ast}
  \Psi_{k \mu^{\prime}}
 [d^{(k)}_{\mu^{\prime} \mu} (\tau ) ]^{2} \; (-1)^{\mu } \;
e^{2 i (\mu - \mu^{\prime}) \chi }\, ,
\end{equation}
where $d_{\mu \mu^{\prime}}^{(k)} (\tau) $ are the $d$-functions for SU(1,1), which are the hyperbolic counterparts of the Wigner $d$ functions for SU(2)~\cite{Varshalovich:1988ct}; that is,
\begin{equation}
  d_{\mu \mu^{\prime}}^{(k)} ( \tau ) = 
  \langle k, \mu |e^{i \tau  \op{K}_{y}} |k, \mu^{\prime} \rangle \, .
\end{equation}
They can be expressed as~\cite{Vilenkin:1991aa,Ui:1970nh}
\begin{eqnarray}
  d_{\mu \mu^{\prime}}^{(k)} (\tau) & = &
  \left [ \frac{\Gamma (\mu + k) \Gamma (\mu - k +1)} 
{\Gamma (\mu^{\prime} + k) \Gamma (\mu^{\prime} - k +1)} \right ]^{1/2}
\frac{1}{\Gamma (\mu -\mu^{\prime} +1 )} \; 
\pFq{2}{1}{k-\mu^{\prime} , k +\mu}{\mu - \mu^{\prime} +1}
{\tanh^{2} (\tau/2) }  \nonumber \\
& \times & \left [ \cosh (\tau /2) \right ]^{-2k + \mu^{\prime}-\mu}
\left [ \sinh  ( \tau / 2) \right ]^{\mu - \mu^{\prime}} \,,
 \label{eq:lpm}
\end{eqnarray}
where ${}_{2}F_{1}$ is the hypergeometric function~\cite{NIST:DLMF}.
In the final expression \eqref{eq:finalds}, we have made use of the
fact that $  d_{\mu \mu^{\prime}}^{(k)} (\tau)$  are real and 
\begin{equation}
  d_{\mu \mu^{\prime}}^{(k)} (\tau) = (-1)^{\mu^{\prime} - \mu} \,
  d_{\mu^{\prime} \mu}^{(k)} (\tau) \,.
\end{equation}
Equation~\eqref{eq:finalds} is a closed expression for the SU(1,1)
Wigner function we were looking for. Alternatively, one can rewrite
the Wigner kernel  $\op{w} (\zeta)$ in the form 
\begin{equation}
  \op{w} (\zeta) =  \exp \left ( i \pi [
 \op{K}_{0} \cosh \tau  -  \tfrac{1}{2} (e^{i \chi} \op{K}_{+}
      + e^{- i \chi} \op{K}_{-} ) \sinh \tau  ] \right )  \, ,
\end{equation}
which can be disentangled as
\begin{equation}
  \op{w} (\zeta) = e^{\gamma_{-} \op{K}_{-}} \;
  e^{i \pi  \op{K}_{0}} e^{\ln \gamma_{0} \op{K}_{0}}
  e^{\gamma_{+} \op{K}_{+}} \, ,
\end{equation}
with
\begin{equation}
  { \gamma_{\pm} = e^{\pm i \chi} \tanh \tau \, ,
  \qquad
  \gamma_{0} = 1/\cosh^{2}  \tau}  \, .
  \end{equation}
The action of this operator in the basis $\{ |k, \mu \rangle\}$ can be easily found by expanding the exponentials. After a lengthy calculation the final result coincides with \eqref{eq:lpm}.

 %%%%%%%%%%%%%%%%%%%%%%%%
\begin{figure}[t]
  \begin{center}
    \includegraphics[width=.80\columnwidth]{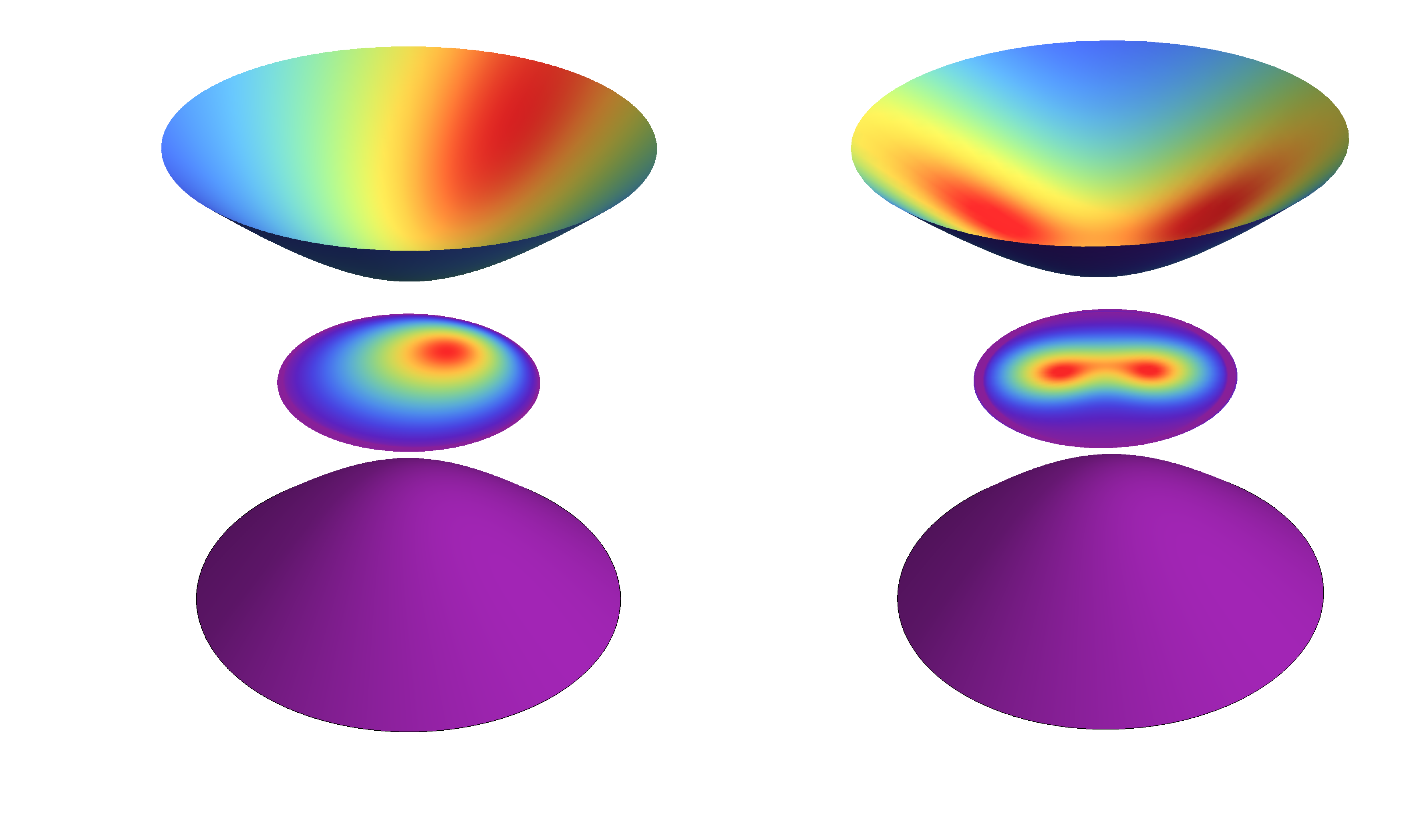}
  \end{center}
  \caption{Density plots of the Wigner function on the upper sheet of
    the hyperboloid $\mathbb{H}_{2}$ and the {associated}
    distribution in the unit disk, obtained from the upper sheet by
    stereographic projection from the south pole. Left panel
    {is} a two-mode squeezed vacuum \eqref{eq:TMSV} with
    $k=1/2$ and $\xi = 0.485$. Right panel is a factorized state \eqref{eq:cohsq} of a coherent  and a single-mode   squeezed state with $\alpha = 1$ 
    and $\xi = 4 + i/2$. {The colormap goes from dark blue (corresponding to 0) to red (corresponding to the maximum value).} }
  \label{fig:TMSV}
\end{figure}
%%%%%%%%%%%%%%%%%%%%

A word of caution is in order here. Strictly speaking, Wigner functions can be properly defined only for a single irrep, where the concept of phase space is uniquely defined. Nonetheless,since the Hilbert space of our original two-mode problem splits as in \eqref{eq:split}, our Wigner function appears as a sum over all the irreps in the discrete positive series.

Let us consider the relevant example of SU(1,1) coherent states,
defined as~\cite{Perelomov:1986kl}
\begin{equation}
  |\xi, k \rangle = \op{S} (\zeta ) |k,k \rangle \,,
\end{equation}
where $\xi = \tanh ( \tau/2) e^{ i \chi}$ and
$\zeta = \tfrac{1}{2} \tau \, e^{i \chi}$. These states live in the
irrep $k$ and for $k=1/2$ they are nothing but two-mode squeezed
vacuum states, which in the photon-number basis read
\begin{equation}
  \label{eq:TMSV}
  |\Psi \rangle = \sqrt{1- |\xi|^{2}} \sum_{n=0}^{\infty}
  \xi^{n} \;  |n\rangle_{a} |n \rangle_{b} \, .
\end{equation}
Using this form in our general formula, we get an involved
expression. The result appears in Fig.~\ref{fig:TMSV}. As we can
appreciate, the squeezing appears here as a displacement (and not
merely as a deformation, as in the case of continuous variables). The
limit of infinite squeezing corresponds to displacing the state to the
infinity in the upper sheet, which means that the function tends to
the border of the unit disk. Note that the metric in the unit
disk is not Euclidean, but hyperbolic. This means, that as we approach
the boundary, big squeezing translates in small displacements.

As a second example, we consider the factorized state
\begin{equation}
  \label{eq:cohsq}
  |\Psi \rangle = | \alpha \rangle_{a} | \xi \rangle_{b} \, ,
\end{equation}
where $| \alpha \rangle_{a} $ is a coherent state in mode $a$ and
$| \xi \rangle_{b} $ a single-mode squeezed state in mode $b$. The
decomposition \eqref{eq:finalds} now involves a sum over all
irreps. This sum can be split into integer and half-integer values,
which translates in the presence of two peaks in the corresponding
Wigner function. The displacement from the origin of these peaks is
related to the squeezing, as before, and can be appreciated in
Fig.~\ref{fig:TMSV}. 

\section{Application to an SU(1, 1) interferometer}

\begin{figure}[t]
  \begin{center}
    \includegraphics[width=\columnwidth]{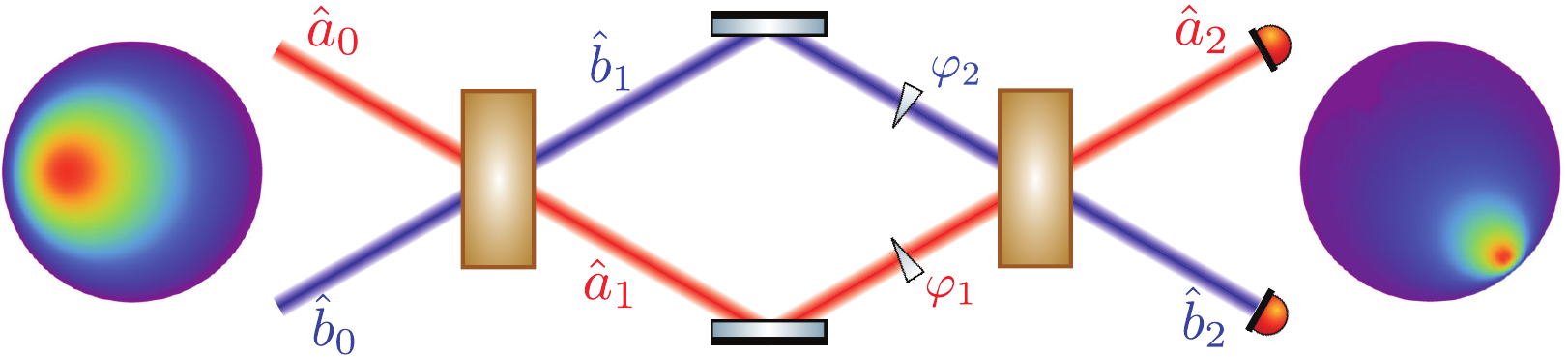}
  \end{center}
  \caption{Schematic diagram of an SU(1,1) interferometer, which consists in a Mach-Zehnder configuration in which the beam splitters have been replaced with optical parametric amplifiers (OPAs). Each arm of the interferometer undergoes a different phase shift. On the left we show the Wigner function on the unit disk corresponding to an input state that is a two-mode squeezed vacuum \eqref{eq:TMSV}, with $k=1/2$ and $\xi = 0.5$, whereas on the right we see the corresponding Wigner function for the output. We have assumed a gain $\mathcal{G}=0.5$ and $\Phi = \pi/2$.}
  \label{fig:setup}
\end{figure}

{To check how the Wigner formalism developed thus far works, let us consider a typical SU(1,1) interferometer, as sketched in Fig.~\ref{fig:setup}.  Two input modes interact via an optical parametric amplifier (OPA) with gain parameter $\mathcal{G}$. The action of the OPA is given by the squeeze operator $\op{S} (\zeta)$, defined in Eq.~\eqref{eq:squeezeop}, with $\zeta = \mathcal{G} e^{i \vartheta}$. After the first OPA, the upper path undergoes a phase shift $\phi_{1}$ and the lower path undergoes a phase shift $\phi_{2}$.  We assume a balanced configuration, where the two OPAs have a fixed phase difference of $\pi$ and the same gain factor. In consequence, the action of the interferometer, which transforms input modes (labeled 0) into output modes (labeled 2), can be concisely expressed as}
\begin{equation}
\hat{T}= \op{S} (\zeta) \; 
e^{i ( \phi_{1} \op{a}^{\dagger} \op{a} 
+ \phi_{2} \op{b}^{\dagger} \op{b})} \; 
\op{S}^{\dagger}(\zeta)  \, .
\end{equation}
{The important observation is that, up to a constant phase, in each subspace with a fixed difference $\op{a}^{\dagger} \op{a} - \op{b}^{\dagger} \op{b}$, this can be written in an compact SU(1,1) notation; viz}
\begin{equation}
{\hat{T}  = \op{S} (\zeta) \; e^{i \Phi \op{K}_{0}} \; 
\op{S}^{\dagger} (\zeta)} \, ,
\end{equation}
with $\Phi = \phi _{a} + \phi _{b}$.

The action of $\op{T}$ (and, hence, of the interferometer) on the input state is $\op{\varrho}_{\mathrm{out}} = \op{T} \, \op{\varrho}_{\mathrm{in}}  \; \op{T}^{\dagger}$ and on the Wigner function reads
\begin{equation}
{W_{\mathrm{out}} (\zeta ) =  
\Tr [ \op{\varrho}_{\mathrm{in}} \; \op{T} \op{w} (\zeta) \op{T}^{\dagger} ] = 
\Tr [ \op{\varrho}_{\mathrm{in}} \, \op{w} (g^{-1} \zeta) ] = 
W_{\mathrm{in}} ( g^{-1} \zeta) \, , }
\end{equation}
where $g$ is the group element corresponding to $\op{T} $, which has the form
\begin{equation}
\label{eq:gsu(11)}
g =\left( 
\begin{array}{cc}
\cos (\Phi/2) +i \sin (\Phi /2) \cosh \tau & i e^{i \chi} \sin
(\Phi /2) \sinh \tau \\ 
-ie^{-i\chi} \sin (\Phi /2) \sinh \tau & \cos (\Phi /2) -i \sin (\Phi /2) 
\cosh \tau 
\end{array}
\right )  \, .
\end{equation}
The action is via M\"obius transformations~\cite{Perelomov:1986kl}; i.e.,
\begin{equation}
g^{-1} \zeta = 
\frac{-\alpha ^{\ast }\zeta +\beta }{\beta ^{\ast }\zeta -\alpha }  \, ,
\end{equation}
where $\alpha$ and $\beta$ are the matrix elements of $g$ in \eqref{eq:gsu(11)}
\begin{equation}
g=\left( 
\begin{array}{cc}
\alpha  & \beta  \\ 
\beta ^{\ast } & \alpha ^{\ast }%
\end{array}%
\right) ,\qquad |\alpha |^{2}-|\beta |^{2}=1 \, .
\end{equation}
{We thus have a closed formula that allows one to compute the Wigner function of the output state for any input state in the interferometer.}

{As an example, we show the action of an interferometer with gain $\mathcal{G}=0.5$ and $\Phi = \pi/2$, on an input state that consists of a two-mode squeezed vacuum \eqref{eq:TMSV}, with $k=1/2$ and $\xi = 0.5$. The interferometer action can be clearly identifies: it squeezes and the rotates the state. This phase-space approach allows one to understand this action in a very clear and intuitive way.}

%%%%%%%%%%%%%%%%%%%%%%%%
%%%%%%%%%%%%%%%%%%%%

\bigskip

\section{Concluding remarks}

Quantum phenomena must be depicted in the proper phase space. This is unanimously recognized for continuous variables (with the complex plane as phase space), for spinlike systems (with the Bloch sphere as the underlying manifold), for orbital angular momentum (represented in the cylinder), {and for other systems}. Surprisingly, the physics related to the SU(1,1) symmetry is not displayed on the hyperboloid, the natural arena for these phenomena.

What we {have accomplished} here is to provide a practical framework to represent SU(1,1) states in an appropriate way. Apart from the intrinsic beauty of the formalism, our compelling arguments should convince the community of the benefits that arise using the proper phase-space tools to deal with these systems.

\section*{Acknowledgments}

We acknowledge financial support from the Mexican Consejo Nacional de Ciencia y Tecnologia (CONACYT) (Grant 254127), the Spanish Ministerio de Ciencia e Innovaci\'on (MICINN) (Grant PGC2018-099183-B-I00) and the Natural Sciences and Engineering Research Council (NSERC) of Canada

%\bibliographystyle{unsrtnat}
%\bibliography{angular}

\end{document}